# Nuancing the Neuron

A Review of

**The Memory Process: Neuroscientific and Humanistic Perspectives**
By Suzanne Nalbantian, Paul M. Matthews, and James L. McClelland (Eds.)
Cambridge, MA: MIT Press, 2011. 440 pp; ISBN 978-0-262-01457-1

Reviewed by Liane Gabora and Apara Ranjan

Corresponding Author:
Dr. Liane Gabora
Department of Psychology
University of British Columbia
Okanagan campus, 3333 University Way
Kelowna BC, V1V 1V7, CANADA
Email: liane.gabora@ubc.ca

*The Memory Process*, edited by Suzanne Nalbantian, Paul M. Matthews, and James L. McClelland, is an intriguing and well written book that provides a groundbreaking overview of diverse approaches to understanding memory that sets the agenda for an interdisciplinary approach to the topic.

Memory has long been a focus of investigation and interest in both the sciences and the humanities. The way memory enriches and distorts lived experience has been widely explored in literature and the arts. Our fascination with the subject is increasingly evident in popular culture, with the widespread proliferation of novels and movies in which events play out in a nonlinear fashion that reflects how memories of them are woven together in the minds of the characters involved. Scientific approaches to memory have focused on the study of amnesiacs, neuroimaging studies, and cognitive studies of the formation, retrieval, and forgetting of memories. Until now, however, humanistic and scientific investigations of memory have been carried out independently. This book provides an exemplary illustration of how disciplinary boundaries can be transcended, showing how approaches to memory from the humanities and the sciences can revitalize one another. With 19 chapters written by academics from a range of disciplines including neuroscience, psychology, psychiatry, cognitive science, cultural studies, philosophy, bioethics, history of science, art, theatre, film, and literature, it is undoubtedly one of the most interdisciplinary books on library shelves. Together these diverse perspectives provide a broad and stimulating overview of the memory process.

The book is divided into five sections. The first two sections outline fundamental concepts and ongoing debates in memory research from a scientific perspective. They do an excellent job of explaining basic concepts such as procedural versus declarative memory and phenomena such as the stability, veridicality, and the localization of memory. The first section (chapters 1 to 4), "Scientific Foundations" presents historical scientific perspectives on how memories are encoded, starting with the notion of the engram as an enduring physical change in the brain constituting a fundamental unit of memory, and showing how conceptions of memory have evolved. It discusses molecular genetic approaches to how experiences are consolidated in long term memory, how we derive meaning from remembered experiences by relating them to things we already know, and the role of sleep and dreaming in this process. It includes a fascinating chapter by Changeux on the cellular mechanisms of brain plasticity and its implications for the epigenetic variability of memory. While we point to scientific reasons for not framing the honing and transmission of artistic tools and techniques across generations in a Darwinian framework (Fracchia & Lewontin, 1999; Tempkin & Eldredge, 2007; Gabora, 2008), the author makes the important point that each generation builds on and modifies the memories of preceding generations.

The second section (chapters 5 to 8) is titled "Scientific Phenomena and Functioning." This section presents neuropsychological studies of memory in animals and humans, including neuroimaging and pharmacological studies of memory; the relationship between memory, synaptic connectivity, and emotions; and computational approaches to memory. As long-time fans of the work of James McClelland, we were particularly interested in the chapter on connectionist models of memory. We found it highly interesting, though we note that distributed representation is not inconsistent with the notion that individual neurons respond to specific microfeatures, and that the *context* affects which neurons participate in the distributed encoding of a particular instance of something. This paves the way for a scientific theory of the neural underpinnings of creative insight as the context-driven activation of a new variation on a previous pattern of distributed activation (Gabora, 2010; Rossmann & Fink, 2010). We found the discussions of emotional aspects of memory, such as the dynamic processing of fear memories in the chapter by Ledoux and Doyere, and how implicit and explicit processing systems for emotional memory affect decision making and creativity in the chapter by Rolls, to be particularly informative.

The middle section of the book (chapters 9 to 11) is titled "Crossroads to the Humanities." This section serves to bridge the scientific and humanities perspectives. It address philosophical and ethical issues pertaining to memory including how it sheds light on ontological and epistemological problems, and the role of memory in one's sense of personal identity. An idea addressed in this section, which echoes back to points made in the chapter on the epigenetic variability of memory, is the commonly held but only vaguely understood notion that memory resides at not just the individual level but also the cultural level.

The last two sections present memory from the perspective of the humanities. The fourth section (chapters 12 to 15) is titled "Literary Data for Memory Studies." This section provides intriguing examples of how autobiographical memory is portrayed in literature and discusses what these portrayals imply more generally for how memory works. Of particular interest is the use of metaphors to describe memory phenomena, and the parallels between how memory is portrayed and understood in literature and contemporary neuroscience, such as that it is not replicative but prone to ongoing recategorization and reconstruction. The in-depth introspective accounts of specific cases involving specific individuals give the reader a fascinating glimpse into not just the intricacies of emotional memory and its contextual construction but also the worldviews of modernist authors such as Virginia Woolf, James Joyce, and William Faulkner.

The fifth and final section (chapters 16 to 19) is titled "Manifestations in the Arts." It provides a novel outlook on how human memory is understood and interpreted in a variety of artistic domains. Topics of particular interest in this section are the role of procedural memory and emotional memory in eliciting responses to visual art, and the way movies portray and in some cases are even inspired by the phenomenon of forgetting. We were intrigued by the somewhat baffling notion that strong emotions leave a distinctive trace and that such traces can be found throughout the body. The chapters on music very clearly weave together expert knowledge of both neuroscience and the artistic domain of music, and demonstrate how simultaneous consideration of a topic from different perspectives can reinforce each other resulting in a more well-rounded and nuanced perspective. Of particular interest was the discussion of how music triggers autobiographical memories.

*The Memory Process* will be of interest to non-academics as well as scholars in a variety of fields. The editors [[AU: Did you per chance mean "editors" here instead of authors? yes]] make the valid point that something was gained and something was lost as the scientific study of memory made a shift away from introspective studies, and brain imaging approaches began to predominate. We were convinced that there is a wealth of complementary approaches to memory, and that bridging these approaches can be fruitful. Analyses of how memory has been understood and explored in works of art and literature provide scientists with a treasury of striking and detailed examples that may be difficult to obtain with time constraints in controlled laboratory conditions. Scientists are forced to come face to face with the exciting challenge of explaining, not just findings obtained from studies using simple memory tasks, but also complex, multifaceted questions about the organization of memory, such as how it is portrayed in dramatic works ranging from *Hamlet* to *The Glass Menagerie*. The book paves the way for those in the humanities to incorporate scientific findings, such as evidence for the extent to which memory is a constructive process, into their analyses of texts and artworks. We believe that artists too have much to gain from this effort. Looking at art in part through a scientific lens, the artist can consciously incorporate facts about how artistic stimuli are encoded and retrieved into their artistic works, which may increase the impact on the viewer. We highly recommend this pioneering book to anyone interested in the nature of memory.